# Near 65ps-fwhm in 20mm-LSO TOF-PET by using a time walk correction based on multi-photon stamps: a Monte Carlo study.

Stephan Walrand, Michel Hesse, François Jamar

Nuclear medicine department, Cliniques Universitaires Saint-Luc, Brussels, Belgium

**Abstract**

**Aim:** Investigate whether a time walk assessment derived from crystal light sharing and multi-photon time stamps can improved the TOF resolution.

**Method:** Monte Carlo simulations were performed for SiPM (FBK-NUV-HD) coupled to conventional 4×4 $mm^2$ pixelated LSO (0.2%Ca:Ce ) and also coupled to 8×4 $mm^2$ pixelated LSO in which a triangular optical window allows light sharing between the two 4×4 $mm^2$ sub-pixels depending on the DOI. A time walk estimation was modelled as a DOI dependent weighted sum of the time differences within the photons detection sequence. The DOI dependent weights were optimized in order to get the best TOF resolution.

**Results:** A strong linear DOI dependence of the light sharing between the two 4×4 $mm^2$ sub-pixels was observed. The TOF resolutions for a 4×4×20 $mm^3$ LSO were 99.7, 81.2, 66.4 ps-fwhm without any correction, with DOI correction and with DOI dependent time walk correction, respectively. The time walk correction was derived from the 7 time stamps of the ($2^k$ )$^{th}$ photons in the sequence, with $k \in [0,6]$.

**Conclusion:** photon time walk can be estimated from the photon time detection series and can be advantageously combined with the DOI assessment to improve TOF resolution.

## Introduction

Although still being far to provide an accurate localization of the annihilation, current TOF information already provides several benefits: a signal to noise ratio improvement [Conti. 2006], information redundancy making the reconstruction more robust in presence of inconsistent data such as incoherencies between measured and actual attenuations [Conti 2010, Walrand 2018] and improvement of the condition number of the inverse problem [Llacer 1982].

Continuously improving the TOF resolution is also part of the physicist dream: with a 25ps-fwhm CTR the annihilation will directly be localized in a 4×4×4 $mm^3$ voxel. As a result the tomographic images will only be affected by an uncorrelated Poisson noise which facilitates interpretation by a human observer [Burgess 2011]. Last, reconstructed low activity regions of interest will no longer be impacted by high taking up neighbouring regions.

Derenzo et al. 2014 performed extensive Monte Carlo (MC) simulations of various crystal types and dimensions coupled to several photo-detectors. They were able to derive an accurate, but sophisticated formula predicting the TOF resolution. For L[Y]SO crystals coupled to SIPM, this formula reduces to a simple formula less accurate, but better illustrating how the different parameters impact the TOF resolution [Walrand et al. 2020]:

$$CTR \approx 2.20 \sqrt{\frac{\tau_d \, (\tau_r + \mathrm{d} + 0.95\ \mathrm{J})}{N}} \qquad (1)$$

Where CTR is the coincidence timing resolution fwhm ; $\tau_r$ and $\tau_d$ are the crystal scintillation light rise and decay times, respectively; $d$ is the optical photon time dispersion in the crystal; $J$ is the Jitter fwhm of the photodetector and $N$ the number of detected scintillation photons given by:

$$N = Y\ T\ PDE \qquad (2)$$

Where $Y$ is the crystal scintillation yield, $T$ is the fraction of scintillation photons transported from the photoelectric effect location to the photodetector sensitive area and $PDE$ is the photodetector efficiency.

Eq. 1-2 show that beside the intrinsic crystal and photodetector characteristics, it is also of prime importance to optimize the light transport in the crystal in order to improve the CTR. As all materials have a reflection coefficient lower than one [Janecek 2012], the transport fraction $T$ can be increased by reducing the number of reflections needed for the photons to reach the photodetector. This reduction also narrows the photon walk length spectrum, which results in a reduction of the photon time dispersion $d$.

This explains why Lambertian reflectors [Janecek and Moses 2008], such as Teflon, are often used for the lateral crystal surface. Indeed, except for large DOI, a large part of the scintillation photons will hit the lateral surfaces of pixelated crystals with a small angle of incidence, resulting in many specular reflections before reaching the detector. By allowing reflections along an angle larger than that of the incidence, Lambertian reflector redirects a fraction of these photons towards the photodetector.

It has been shown that improving the lateral surfaces polishing increases the transport fraction $T$ [Huber et al. 1999] and allows CTR down to 100 ps in 20mm long crystal [Gundacker et al. 2020]. However, it was also recently shown that leaving some roughed area on one lateral surface in the small DOI region further improved transport fraction $T$ and photon time dispersion $d$, resulting in a CTR improvement [Berg et al. 2015].

Beside improving the light transport, another important aspect is to correct the photon detection time by its time walk inside the crystal, i.e. the term d in eq. 1. Several methods have been proposed in order to assess the DOI (see [Berg and Cherry. 2018] for a review), afterwards the detection time can be corrected by adding the DOI corrected with an apparent speed of the light in the crystal. This optimized apparent speed is lower than the actual one as a consequence of the photon walk complexity [Walrand et al. 2020]. Photon time walk corrections based on interaction localization in monolithic crystal [Vinke et al. 2010] or based on scintillation pulse shape analysis [Wiener et al. 2010] have also been proposed.

The aim of this study is to investigate by MC simulations whether DOI assessment, using a simple light sharing between crystal pixels, combined with multi-photon time stamps could better predict the photon time walk in order to improve the CTR.

**Material and methods**

*MC simulations*

A previously described and validated MC code [Walrand et al. 2020] was used. The MC code includes a polish parameter: polish = 0.60 corresponds to a standard mechanical polishing while polish = 0.95 corresponds to a fine chemical etching. The code was previously validated versus the scintillation photon counts experiments of [Huber et al. 1999] and [Bauer et al. 2009] for different crystal polishing levels, and also versus the CTR experiments of [Degenhardt et al. 2009, 2012] which used the Philips LYSO-SiPM block.

Present simulations were performed for a LSO 0.2%Ca:Ce crystal coupled to a FBK-NUV-HD SiPM. This setup already achieved a CTR of 58 and 98 ps-fwhm for a 3 and 20 mm-long thoroughly polished crystal, respectively [Gundacker et al. 2020]. First the MC code was validated for these two setups. Afterwards, conventional and light sharing LSO-SiPM blocks of various lengths were investigated. All the crystal were assumed chemically etched, i.e. polish = 0.95. Fifty thousand pairs of coincident photelectric effects were simulated for all setups. The photoelectric effect DOI was randomly drawn according to the attenuation coefficient of the crystal. Crystal and SiPM characteristics were extracted from [Gundacker et al. 2020].

*DOI assessment*

A previous light sharing crystal block design allowed a 70 ps-fwhm delta improvement of the CTR using standard polished LYSO crystals [Walrand et al. 2020]. This design was here improved: the light sharing was limited within crystal pixel pairs rather than between 4 adjacent crystal pixels in order to reduce the counts reductions resulting from the splitting between the pixels, and the rectangular optical window was replaced by a triangular one in order to magnify the light sharing dependence on the DOI.

Figure 1 shows the novel light sharing crystal 2-pixel block. A tilted partial saw cut is performed in the middle of a 4x8xL $mm^3$ crystal, the crystal being totally cut at the SiPM interface and not at all in the front crystal surface. The crystal is covered by 4 Teflon layers, but the partial saw cut is left empty in order to allow total lossless reflection for incidence angle larger than arcsin(1/n) where n is the index of refraction of the crystal, i.e. ≈ 1.8 for L[Y]SO crystal. With this design, the probability that a photon can enter into the lateral pixel decreases when the DOI increases and, as a result, also does the primary (P) to lateral (L) counts ratio (P/L). The crystal is assumed to be chemically etched during 15 min after the partial sawing.

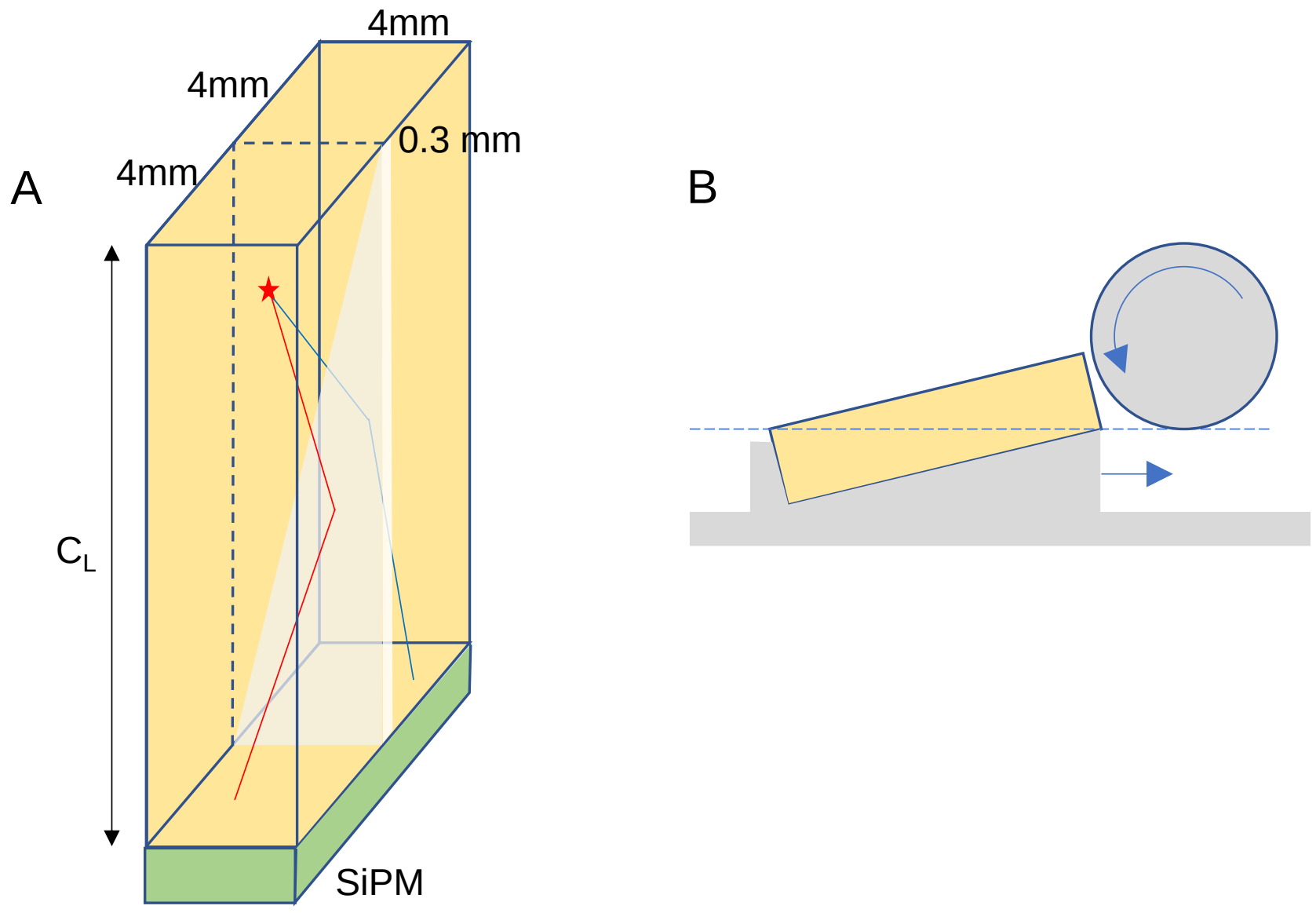

Figure 1: A: Open view of the novel light sharing crystal block design. Transparent white-grey: empty triangular 0.3 mm-width saw cut shaping a complementary triangular optical window between the two 4×4 $mm^2$ crystal pixels. The red curve represents one photon undergoing a total lossless reflection on the crystal-saw interface and thus remaining in the primary pixel, while the blue one represents a photon entering into the lateral pixel via the optical window and undergoing a reflection on the back side covered by 4 Teflon layers. B: the triangular saw cut can easily be machined by translating a 4×8×$C_L$ $mm^3$ crystal under a precision circular saw in the same way that for the other sides, but with the crystal lying on a tilted holder.

*Time walk estimation*

We hypothesize that for a specific DOI, or more exactly for a specific P/L ratio, the detection time series of the scintillation photons contains some information about the walk length. According to this assumption the assessed TOF was modelled as:

$$TOF = t_1 + A_0(P/L) + \sum_{p=1}^{N} A_p(P/L)\ \left(t_{p+1} - t_1\right) - t_{annih} \qquad (3)$$

where *P/L* is the primary to lateral counts ratio measured for the considered annihilation, $t_p$ is the detection time of the photon *p* without considering in which pixel this photon is counted, *N* is the number of photon used for the assessment, $t_{annih}$ is the annihilation time. The $A_p$ parameters have to be optimized in order to get the best CTR. Note that as only the TOF difference between the two detected 511 keV gamma rays is used in the reconstruction process, nor the time origin nor the annihilation time do matter.

The third term of eq. 3 models the time walk correction to be applied on the first detected photon time. Intuitively, when one or a few scout photons are detected very promptly after the photoelectric effect in an uncommon way, their time differences with the photons troop are likely larger. Thus, in this case the third term is expected to delay the detection time, obviously if the coefficients $A_p$ are mainly positive.

Note that when N = 0, eq. 3 reduces to:

$$TOF = t_1 + A_0(P/L) - t_{annih} \quad (4)$$

Eq. 4 is just a simple optimized DOI correction applied on the time stamp of the first detected photon: detection time of large P/L, i.e. large DOI, are delayed if $A_0$ is positive.

To reduce the number of photon times to stamp, the following assessment was also evaluated:

$$TOF = t_1 + B_0(P/L) + \sum_{k=1}^{n} B_k(P/L)\left(t_{2^k} - t_1\right) - t_{annih} \quad (5)$$

The exponential progression takes benefit from the fact that during the scintillation pulse detection the successive photon detection times are closer and closer when approaching the pulse maximum, which thus does not require a fine sampling.

In order to obtain a matrix optimization problem, the ratio *P/L* was discretized within its range in 10 equidistant levels $l$. The corresponding parameters set $\{A_p^l\}$ and $\{B_k^l\}$ were optimized in order to get the best CTR fwhm computed as 2.35 fold the standard deviation of the simulated TOF differences within pairs of photoelectric effects. As the time origin vanishes in the TOF difference, one can assume $A_0^0 = 0$ and $B_0^0 = 0$ to avoid singular system matrix.

## *Results*

### *MC code validation*

Table 1 shows the validation of the MC code versus experiments performed with LSO 0.2%Ca:Ce FBK-NUV-HD block [Gundacker et al. 2020]. The best agreement was obtained for a polish coefficient of 0.94 corresponding to chemical etching during about 15 min.

Table 1: experimental and MC simulated CTR.

| crystal shape | exp. CTR [ps-fwhm] | MC CTR [ps-fwhm] |
|---|---|---|
| 2x2x3 mm3 | 58 | 55 |
| 2x2x20 mm3 | 98 | 99 |

### *DOI assessment*

Fig. 2A shows that the primary counts are always higher than the lateral one, which allows the determination of the pixel in which the photoelectric effect occurred. Fig. 2B shows that there is a bijective relation between the primary to lateral counts ratio P/L and the DOI.

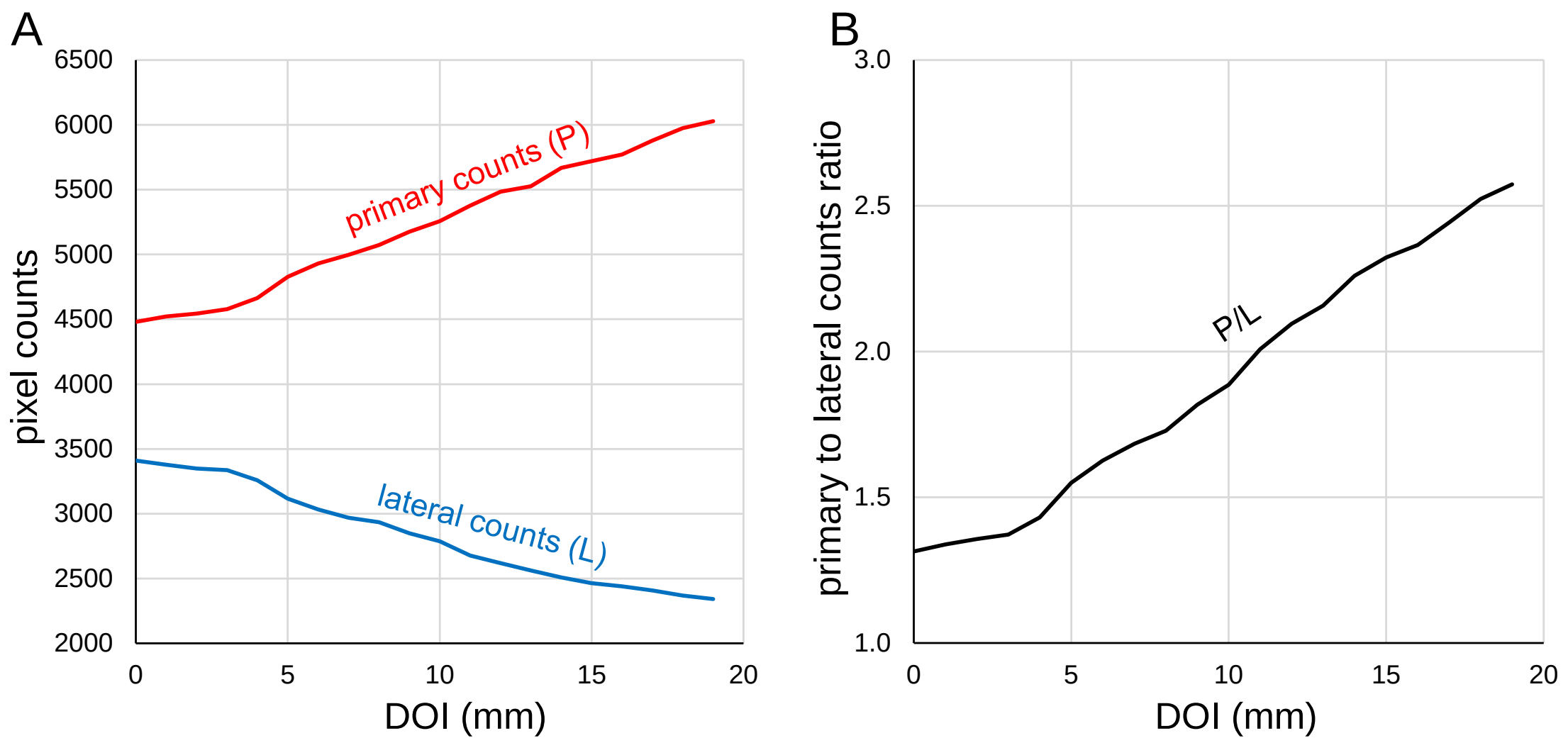

Figure 2: A: primary and lateral counts and B: primary to lateral counts ratio, both as a function of the DOI for a 4×4×20 mm$^3$ LSO 0.2%Ca:Ce - FBK-NUV-HD block.

*CTR assessment*

Figure 3 shows the CTR improvement obtained using eq. 3 as a function of the number of photon time stamps used. A first 18ps-fwhm improvement is provided by the DOI correction alone (eq. 4), followed by a 15ps-fwhm improvement using 32 photon time stamps.

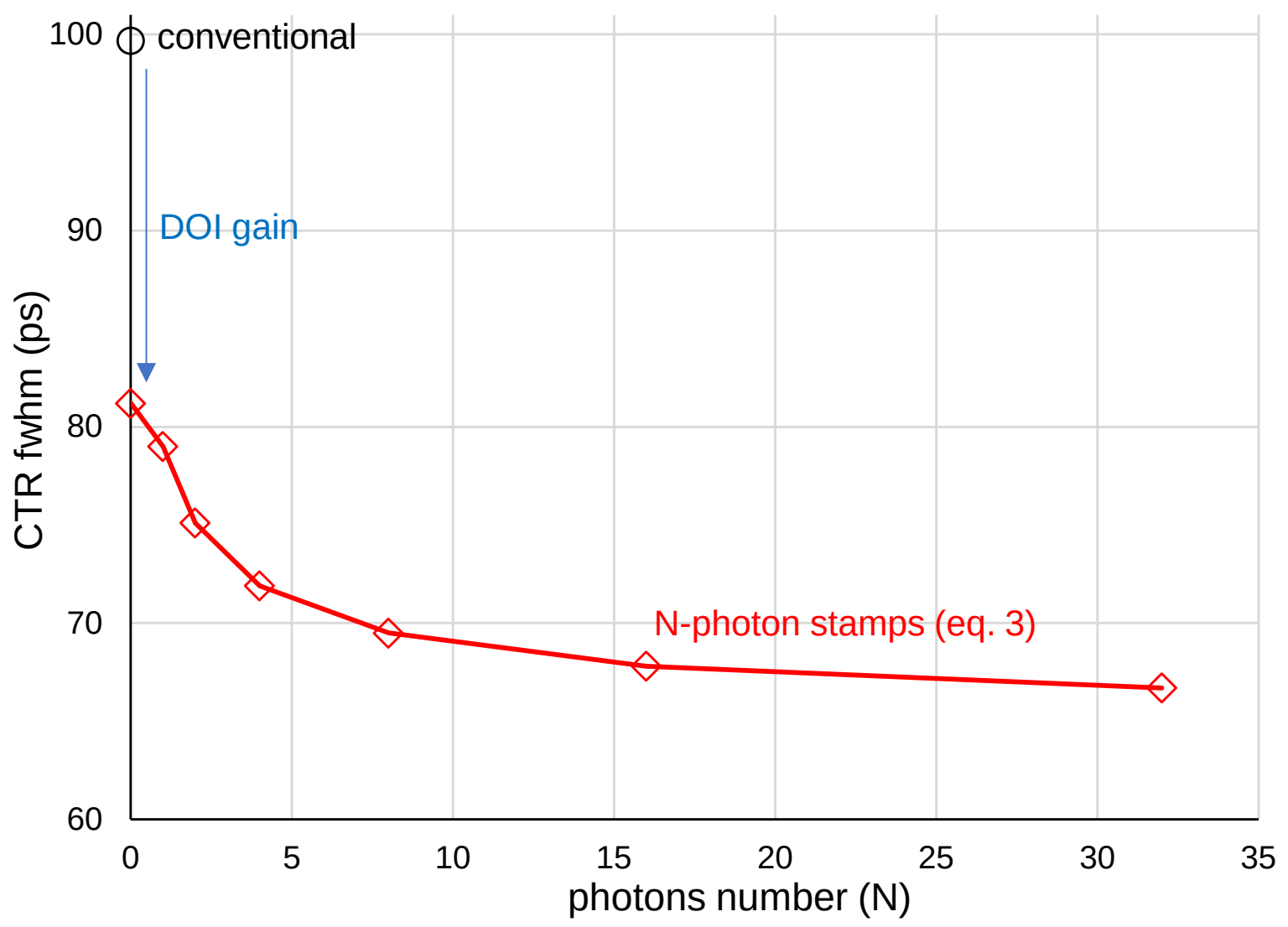

Figure 3: CTR for a 4×4×20 mm$^3$ LSO 0.2%Ca:Ce - FBK-NUV-HD block. Black circle: conventional block setup. Diamonds: light sharing setup (see Fig. 1) using eq. 3 for different number of photon time stamps N. N=0 corresponds to a simple DOI correction (eq. 4).

Fig. 4 shows that eq. 5 provided similar CTR assessments than using eq. 3 but using only $log_2(N) + 1$ time stamps in place of *N+1*. CTR reached 66.8ps-fwhm for n=6 requiring 7 photon time stamps to marginally improved up to 66.4 ps-fwhm for n = 12.

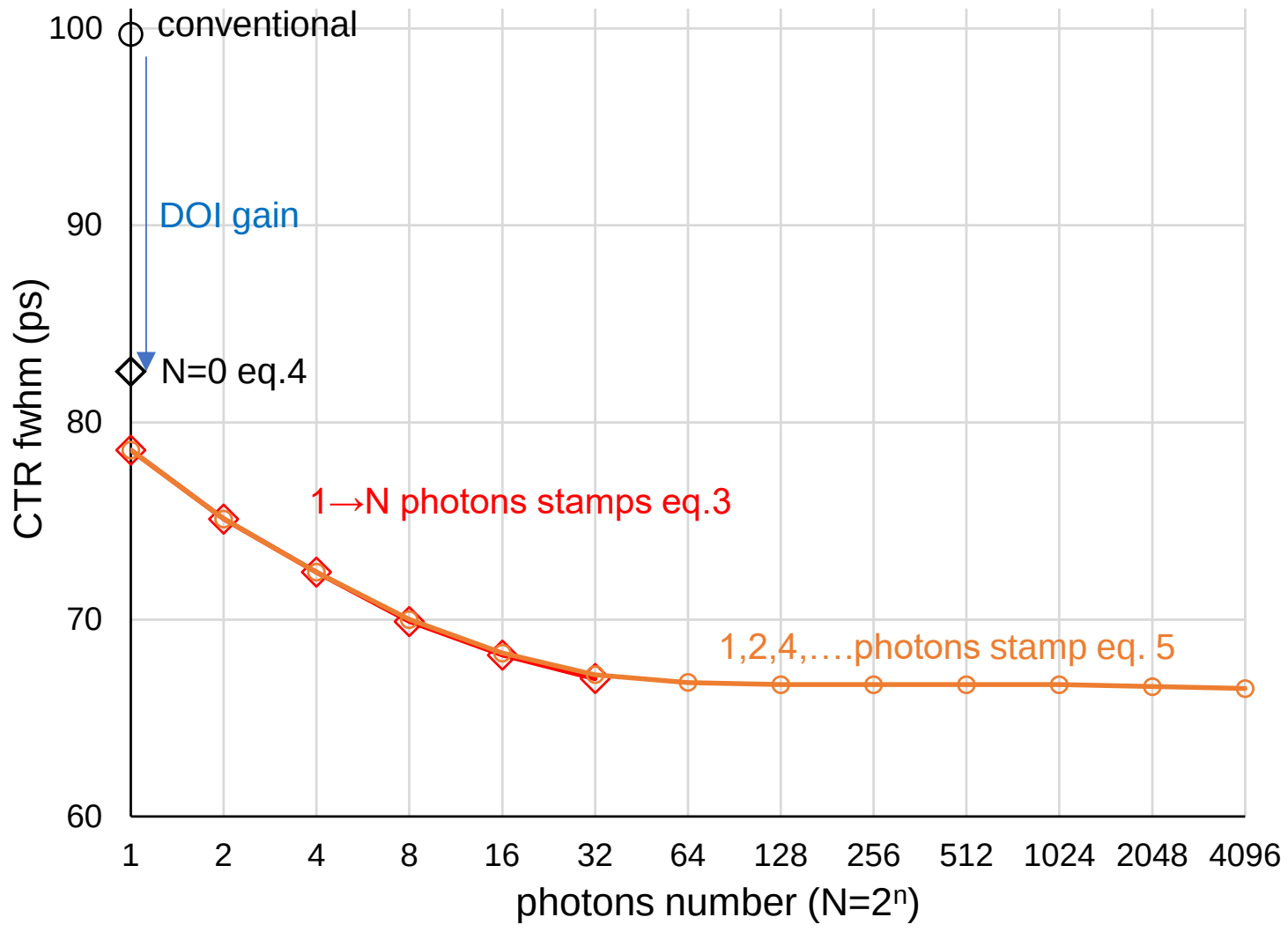

Figure 4: CTR for a 4×4×20 $mm^3$ LSO 0.2%Ca:Ce - FBK-NUV-HD block. Black circle and diamond: without and with simple DOI correction (eq. 4), respectively (symbols were artificially set on the vertical axis, as 0 cannot be represented in log scale). Red diamonds and orange circles: using DOI & time walk correction with eq. 3 using N+1 time stamps and eq. 5 using n+1 time stamps, respectively.

Fig. 5 shows the obtained CTR improvement as a function of the crystal length. DOI correction alone did not provide any improvement for crystal length lower than 7 mm. In contrary, the time walk correction already provided an improvement for a 1mm-length crystal, even resulting for crystal length shorter than 3 mm in a CTR better than what could be expected after an exact optical time walk correction (see discussion).

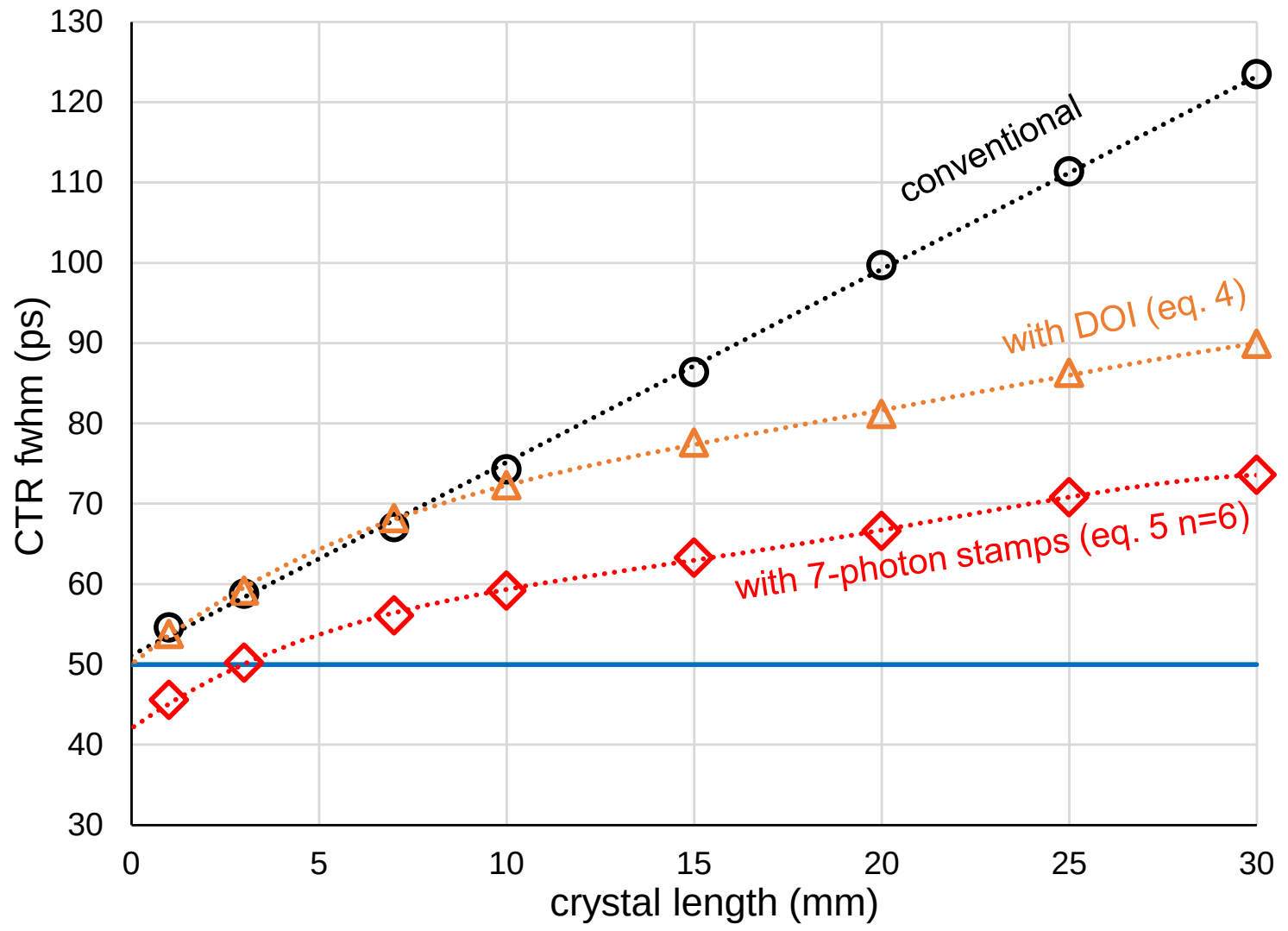

Figure 5: CTR as a function of the crystal length for a 4×4×$C_L$ $mm^3$ LSO 0.2%Ca:Ce - FBK-NUV-HD block. Black circles: conventional block setup. Orange triangles: using simple DOI correction (eq. 4). Red diamonds: using DOI & time walk correction (eq. 5 with n=6). Blue line is the extrapolation of experiment to $C_L = 0$, i.e. the theoretical limit for an exact optical time walk correction, i.e. d = 0.

Fig. 6 shows the optimized $B_k^l$corresponding to the point $2^6$ in fig. 4. For low DOI the ponderation $B_k^l$ exhibits a convex shape, which further moves to a concave shape for larger DOI.

$B_k^l$

| $2^k$ \ P/L level l | 0 | 1 | 2 | 3 | 4 | 5 | 6 | 7 | 8 | 9 |
|---|---|---|---|---|---|---|---|---|---|---|
| 2 | 0.12 | 0.12 | 0.13 | 0.19 | 0.16 | 0.21 | 0.15 | 0.24 | 0.22 | 0.15 |
| 4 | 0.09 | 0.14 | 0.12 | 0.10 | 0.13 | 0.10 | 0.16 | 0.10 | 0.14 | 0.14 |
| 8 | 0.23 | 0.15 | 0.17 | 0.06 | 0.07 | 0.10 | 0.14 | 0.16 | 0.20 | 0.19 |
| 16 | 0.16 | 0.13 | 0.17 | 0.13 | 0.07 | 0.05 | 0.01 | 0.07 | 0.05 | 0.02 |
| 32 | 0.08 | 0.13 | 0.09 | 0.14 | 0.12 | 0.09 | 0.04 | -0.03 | 0.00 | 0.06 |
| 64 | 0.13 | 0.12 | 0.10 | 0.17 | 0.18 | 0.19 | 0.22 | 0.19 | 0.13 | 0.12 |

Figure 6: Colour scale representation of the optimized $B_k^l$ parameters corresponding to the point N=$2^6$ in fig. 4.

## Discussion

The present MC study showed that the DOI correction derived from the light sharing between two crystals allowed a 18ps-fwhm improvement resulting in a 82ps-fwhm CTR for a thoroughly polished 4×4×20 mm$^3$ LSO crystal. This light sharing is performed by simply leaving empty a triangular saw cut machined in the middle of a 8×4×20 mm$^3$ crystal.

Partial saw cuts is a well-known technique already used in the eighties for the pioneer commercial Exact HR PET: the goal at that time being to implement a light sharing between the 4 PMTs depending on the hit crystal pixel. In addition to simplify the reflective covering process, leaving empty the saw cut also provides total lossless reflections for incidence angles larger than 34 degrees in LSO crystal.

The study also showed that the proposed simple time walk assessment, based on 7-photon time stamps, provided an additional 15ps-fwhm improvement leading to a CTR close to 65ps-fwhm for the 20mm-length crystal.

The fact that the method already provided a 8ps-fwhm improvement for vanishing crystal length (fig. 5) shows that the method is not only a time correction of the optical walk, but of the total walk involved during the scintillation process, e.g. also including the electron-hole pair ancestor walk to a crystal luminescent centre (see [Kozhik et al. 2020] for a detailed description of the scintillation process in crystal). Obviously this estimation is not perfect, but fig. 5 clearly shows that its accuracy is sufficient to obtain a significant CTR improvement on the chole crystal length range.

This total walk associated to the first detected photon is estimated using its detection time differences versus the successive $(2^k)^{th}$ photons, where $k \in [1,6]$. Indeed, larger are these differences, likelier the detected photon and its ancestors cumulated short walks. In this case its detection time is thus accordingly delayed in order to be closer to detection times of photons descended from a more common history.

In case of using conventional signal thresholding rather than direct photon counting trigger, the time stamps do no need to exactly match the exponential $(2^k)^{th}$ photons series. Indeed, the optimisation of the $B_k^l$ parameters, that can be performed using acquired data of a point source, will implicitly take into account the actual progression.

The implemented time walk correction was intuitively modelled as a weighted sum of the successive photon detection times. More sophisticated expressions based on statistical theory could be investigated. Likely more promising, development of deep learning algorithms for estimating the time walk from the photon detections time sequence should be considered. Moreover, CTR experiment should be carried out to validate the propose concept.

**Conclusion**

Photon time walk can be estimated from the photon time detection series and can be advantageously combined with the DOI assessment to improve TOF resolution. A significant CTR improvement down to 65 ps using 20 mm length LSO crystal can be obtained using a simple light sharing design and 7-photon time stamps recording.